\documentclass[prl,twocolumn,showpacs,nofootinbib,superscriptaddress]{revtex4}
\usepackage[dvips]{graphicx}
\usepackage{color,array,dcolumn}
\usepackage{amsmath}
\usepackage{amssymb}

\begin{document}

\title{Spin-orbit excitations of quantum wells}
\author{A. Ambrosetti}
\email{ambrosetti@science.unitn.it}
\affiliation{Dipartimento di Fisica, University of Trento, via Sommarive 14, I--38050, 
Povo, Trento, Italy}
\affiliation{INFN, Gruppo Collegato di Trento, Trento, Italy}
\author{JM. Escartin}
\email{escartin@ecm.ub.es}
\affiliation{Univ Barcelona, Fac Fis, Dept ECM, Diagonal 647, E-08028 Barcelona, Spain}
\author{E. Lipparini}
\email{lipparin@science.unitn.it}
\affiliation{Dipartimento di Fisica, University of Trento, via Sommarive 14, I--38050, 
Povo, Trento, Italy}
\affiliation{INFN, Gruppo Collegato di Trento, Trento, Italy}
\author{F.Pederiva}
\email{pederiva@science.unitn.it}
\affiliation{Dipartimento di Fisica, University of Trento, via Sommarive 14, I--38050, 
Povo, Trento, Italy}
\affiliation{INFN, Gruppo Collegato di Trento, Trento, Italy}

\begin{abstract}
\date{\today}
Confinement asymmetry effects on the photoabsorption of a quantum well are discussed by means of a
sum-rules approach using a Hamiltonian including a Rashba spin-orbt coupling.
We show that while the strength of the excitation is zero when the spin-orbit coupling 
is neglected, the inclusion of the spin-orbit interaction gives rise to a non zero strength 
and mean excitation energy in the far-infrared region. A simple expression 
for these quantities up to the second order in the Rashba parameter was derived. The effect of
two-body Coulomb interaction is then studied by means of a Quantum Monte Carlo calculation,
showing that electron-electron correlations induce only a small deviation from the
independent particle model result.
\end{abstract}

\maketitle

%\section{Introduction}

Quantum wells, i.e. strongly confined electrons at a semiconducting heterostructure,
are intrinsically very complex systems, in which the local arrangement of
the electronic structure can in principle give rise to a large variety of interesting
subtle phenomena. However, the very strong confinement and the weak potential felt by
the electrons allowed for successfully using simplified Hamiltonians in describing
the phenomenology of these systems.  The most celebrated is the two dimensional
electron gas (2DEG) in the effective mass  and dielectric constant  approximation, and its
numerous variants. At the semiconductor junction, the effective potential is never symmetric
with respect to the junction plane. This implies that electrons feel on average the effect
of an electric field transverse to the plane. In the frame of effective Hamiltonians, this
fact is accounted for by the inclusion of the so-called Rashba term, which simply describes
the interaction between the spin of an electron moving in the plane, and the self induced magnetic
field. For $N$ electrons the 2DEG Hamiltonian reads:
\begin{eqnarray}
\label{ham}
H_{2DEG}=\sum_{i=1}^N \Big(\frac{p^2_x+p^2_y}{2m^*} + \frac{\lambda}{\hbar} (p^{y}\sigma^{x}-p^{x}\sigma^{y})  
\Big)_i + \\ \nonumber
 \frac{e^2}{\epsilon}\sum_{i<j}^N \frac{1}{|\mathbf{r_i}-\mathbf{r_j}|},
\end{eqnarray} 
where $\epsilon$ and $m^*$ are the effective dielectric constant and electron mass in the semiconductor, respectively.
The Rashba interaction \cite{Rashba}:
\begin{eqnarray}
V_{Rashba}=\frac{\lambda}{\hbar} \sum_{i=1}^N [p_i^{y}\sigma_i^{x}-p_i^{x}\sigma_i^{y}] \,, 
\label{rashba}
\end{eqnarray}
where $\mathbf p_i$ is the momentum of the $i-th$ electron, 
and $\sigma_i^x$ and $\sigma_i^y$ are the Pauli matrices acting over 
the spin of particle $i$, has a strength $\lambda$ proportional to the modulus
of the transverse electric field. Given the formal analogy, the Rashba term
is usually reported as a spin-orbit (SO) interaction.
The study of effective SO effects in confined electron nanostructures has been
the object of many experimental and theoretical investigations in the last few years.
The  most interesting fact is that SO links the spin and charge dynamics,
hence opening the possibility of spin control by means of electrical fields. 
It affects the charge transport\cite{chtr,chtr2,chtr3,chtr4,chtr5,chtr6,chtr7} and the
far-infrared absorption in heterostructures\cite{finf,finf2,finf3,finf4,finf5} and can lead to spin inversion\cite{sin,sin2,sin3,sin4,sin5} and 
variations of the electric spin precession in a magnetic field\cite{prec,prec2,prec3,prec4,prec5}. 
It also induces linear magneto-optical Kerr or Faraday effects then allowing
the measurement of the magnetization of materials by means of laser pulse excitation\cite{Kerr}.
In this sense, the Rashba\cite{Rashba} SO contribution has received special attention since it was shown that its 
intensity can be externally tuned via the application of a gate voltage\cite{nitta,nitta2}.
However, it would be very important in view of technical applications
to have a clean and model independent way to measure the initial 
strength of the Rashba SO interaction in the sample, in order to quantitatively control
the induced effects in the physical processes of interest.
 
In this work we show that this measurement can be achieved by 
far infrared spectroscopy  in the quantum well, {\it in absence of external magnetic
fields}. If the transverse electric field, either given by
the confinement asymmetry or external, is absent, then the light absorption is zero. 
On the contrary, when the SO interaction
is present, photoabsorption occurs with a strength and at an energy in the
far infrared domain which are linearly proportional to the transverse electric field strength.
The actual experiment would  provide, in addition, a very stringent test on the
applicability of the Rashba model to real systems.

In the following we will consider  a  quantum well modeled by $N$ electrons laterally 
confined in the $xy$ plane, at density $\rho$, and interacting via the Hamiltonian $H=H_{2DEG}+V_{Rashba}$.
We make use of effective atomic units in order to simplify the notation, defining 
$\hbar=e^2/\epsilon=m^*=1$.
The lenght unit therefore is the effective Bohr radius $a_0^*=a_0 \epsilon/m^*$ and the energy unit the 
effective Hartree $H^*=H \cdot m^*/(m_e\epsilon^2)$. 

The single particle part of Hamiltonian (\ref{ham}) can be solved analitically\cite{Rashba} to give the
following solutions for single particle energies and wave functions:
 \begin{eqnarray}   
\epsilon_{k_{\pm}}={k^2\over2m}\pm\lambda_R |k|~~,
\nonumber
\\
\varphi_{\pm}={1\over\sqrt{2S}}e^{i \mathbf{k}\cdot \mathbf{r}}
\left(\begin{array}{c}1\\\mp i e^{i\psi} \end{array}
\right)\,,
\label{ansol}
\end{eqnarray}
where $\psi=\tan^{-1}{k_y\over k_x}$. Note that the wave function (\ref{ansol})
is an eigenstate of a component of the spin which is perpendicular to the
direction of $\mathbf{k}$. 
In the following we will address  these states as {\it quasi-up} and {\it quasi-down} single particle states
respectively.
For a given chemical potential  $\mu>0$, the two branches of the  excitation energy $\epsilon_{k_{\pm}}$ , 
are filled up to the Fermi momenta $k^F_+$ and $k^F_-$,
where 
\begin{equation}
k^F_{\pm}=\sqrt{2\pi\rho(1\mp\xi)}~~~~~\xi=\sqrt{2/\pi\rho}\lambda\Big(1-\frac{1}{4}\frac{\lambda^2}{\pi\rho}\Big)
\label{oper}
\end{equation}
(see for example Ref.\onlinecite{Lip08}).

%\subsection{Interaction Hamiltonian and excitation operators}

We now turn to analyze the transitions induced
in the quantum well by the interaction with a
linearly polarized (along $y$) electromagnetic wave
propagating along the $x$-direction, i.e., in
the plane of motion of the electrons. The corresponding
vector potential is ${\bf A}(t)=A\sin\theta
\hat j$, with $\theta=\omega t - q x$. The
interaction Hamiltonian  ${\bf J}\cdot{\bf A}/c +
g^* \mu_B\, {\bf s} \cdot (\nabla \times {\bf A})$, where
${\bf J}={\bf v}$, reads
\begin{equation}  
\label{hint}
h_{int}=\sum_i\left({A\over c}v^y \sin\theta 
-{1\over2}g^*\mu_B q A\sigma^z\cos\theta \right)_i
\;\;\;\; ,
\end{equation}
and the velocity operator $v_y$ is
defined as $v^y\equiv -i[ y, H] =
p^y+ \lambda \sigma^x$.

This interaction Hamiltonian yields three excitation operators:
$\sum_i p_i^y$, $\lambda\sum_i \sigma_i^x$ and $\sum_i \sigma_i^z$. The first one 
does not excite the electrons of the system since it commutes with the
quantum well Hamiltonian. The second operator is sub-leading in $\lambda$
with respect to the third operator, and is normally negligible. We are then left
with the excitations induced in the target by the operator
\begin{equation}
S^z={1\over2} \sum_i \sigma_i^z~.
\label{oper2}
\end{equation}
Note that for a different geometry, {\it e.g.} the one corresponding to transitions induced 
in the system by the interaction with a left-circular polarized electromagnetic wave
propagating along the $z$-direction perpendicular to the plane of motion of the electrons,
the relevant excitation operators are 
\begin{equation}
S^{\pm}={1\over2} \sum_i \sigma_i^{\pm}~,
\label{oper3}
\end{equation}
where $\sigma^{\pm}=\sigma_x\pm i\sigma_y$.
For time-reversal invariant systems, like the ones considered here 
for which $\langle 0|S^z| 0\rangle=0$, it is possible 
to show\cite{LS} that the 
strengths and mean excitations energies corresponding
to the operators (\ref{oper3})  coincide with the ones induced by the operator (\ref{oper2}).    
Photoabsorption experiments attempting to measure the strength $\lambda$ of the SO interaction
could then be performed with electromagnetic waves propagating in a direction
either parallel or perpendicular to the quantum well.

%\subsection{Sum Rules}

The strength and mean excitation energy of the transitions induced in
the quantum well by the operator(\ref{oper2}) can be studied in a convenient way by means of 
the  moments $m_k$ of the dynamic form factor $S(S^z,\omega)$ (which is defined as $\sum_n |<0|S_z|n>|^2 \delta(\omega-\omega_n)$):
\begin{eqnarray}  
m_{k}=\int_0^{\infty} d\omega \omega ^{k}S(S^z, \omega)=\sum _{n}\omega ^{k}_{no}
|\langle 0| S^z| n\rangle| ^{2}~,
\label{mom}
\end{eqnarray}
where $\omega_{no}=E_n-E_0$ is the excitation energy.

By using the completeness relation $\sum_n| n\rangle\langle n| =1$ and
the equation $H| n\rangle =E_{n}| n\rangle$, it is possible to write
the moments (\ref{mom}) as the mean values on the ground
state of commutators $[~]$ and anticommutators $\{~\}$ of the
excitation operator $S^z$ and of the Hamiltonian $H$. For
time-reversal-invariant systems, the following sum rules are
derived for the operator $S^z$:
\begin{eqnarray}       
\begin{array}{rcl}
m_{1}&=&\displaystyle\frac{1}{2}\langle 0|[S^z,[ H,S^z]]| 0\rangle\,,\\[10pt]
m_{0}&=&\frac{1}{2}\displaystyle\langle 0|\{S^z,S^z\}|0\rangle\,,\\[10pt]
m_{-1}&=&\displaystyle\frac{1}{2}\langle 0|[[X,H],X ]| 0\rangle\,,\end{array}
\label{sumr}
\end{eqnarray}
where the operator $X$ is the solution of the following equation
\begin{eqnarray}      
[H,X]=S^z~~.
\end{eqnarray}

The moments $m_k$ can be used to define different mean excitation energies, namely
\begin{eqnarray}
E_{1,-1}=\sqrt{m_1/m_{-1}}
\end{eqnarray}
and
\begin{eqnarray}
E_{1,0}=m_1/m_{0}~,
\label{e10}
\end{eqnarray}
which satisfy the relations
\begin{eqnarray}
E_{1,0}\ge E_{1,-1}~.
\label{e1-1}
\end{eqnarray}
These energies strictly coincide only when the whole  strength is exhausted by a single state. 
However, in presence of collective states, they can be used to study the spreading of the
excitation strength\cite{LS} and to bind the energy of the collective modes.

%\section{Results}

%\subsection{Independent particle model}

In the single particle picture and to the lower order in $\lambda$, the momenta $m_{1}$,$m_{0}$, and 
$m_{-1}$ can be computed analytically. The expressions are: 
\begin{eqnarray}
\begin{array}{rcl}
m_{1}&=&-\frac{1}{2}\langle 0|V_{Rashba}| 0\rangle=N\lambda^2\Big(1-\frac{1}{3}\frac{\lambda^2}{\pi\rho}\Big)\,,\\[10pt]
m_{0}&=&{N\lambda\over\sqrt{8\pi\rho}}\Big(1-\frac{1}{4} \frac{\lambda^2}{\pi\rho}\Big) \,,\\[10pt]
m_{-1}&=&\frac{S}{8\pi}  \,.\end{array}
\label{stren}
\end{eqnarray}
 The corresponding mean excitation energies (\ref{e10}) and (\ref{e1-1}) become:
\begin{eqnarray}
\begin{array}{rcl}
E_{1,0}&=&=\sqrt{8\pi\rho} \lambda\Big(1-\frac{1}{12}\frac{\lambda^2}{\pi\rho}\Big)\,,\\[10pt]
E_{1,-1}&=&=\sqrt{8\pi\rho} \lambda\Big(1-\frac{1}{6}\frac{\lambda^2}{\pi\rho}\Big)\,.
\label{ener}
\end{array}
\end{eqnarray}
Some comments are in order here. The total strength $m_0$ and the energies (\ref{ener}) vanish
for $\lambda\to 0$. As anticipated in the introduction, light absorption in the system only occurs
in presence of SO interaction and the excitation strength and energy are {\it linear} in $\lambda$
at the main order. This fact characterizes this excitations as a genuine effect of the well asimmetry,
or of the presence of a transverse external magnetic field described by the Rashba SO term.

At the lower order in $\lambda$ the two energies (\ref{ener}) coincide, and the higher order
corrections are close each other and very small (of order $\lambda^2$). As a consequence the
excitation strength is practically concentrated in a single peak. This peak is measurable 
since it lies in the far infrared region. In fact taking,  as an example, a GaAs quantum well
for which  $\epsilon=12.4$ and $m^*=0.067m_e$, yelding $H^*=11.86meV$ and 
$a^*_0=97.93 \mathring{A} $, one gets for $E_{1,0}$ a value ranging from 2 to 6 meV, for densities
in the range $(2--13)10^{11}cm^{-2}$ and $\lambda\simeq 10^{-9}eV cm\simeq 0.1a.u. $. 

We want to stress the fact that the excitation is linear in $\lambda$ only in absence of an external
magnetic field, which would immediately make such dependence quadratic, therefore strongly affecting
the possibility of accurately measure the SO strength.
We also stress that, since the excitation operator $S_z$
commutes with the Coulomb interaction, the result $m_1=-\frac{1}{2}\langle 0|V_{Rashba}| 0\rangle$ 
for the energy weighted
sum rule is $exact$ for the total Hamiltonian (\ref{ham}). Deviations from the result (\ref{stren})
for $m_1$ are due to two-body correlations induced in the ground state by the Coulomb interaction.
Finally we note that result (\ref{stren}) for $m_{-1}$, valid at all orders in $\lambda$, is exact
for the single particle model where the Coulomb interaction is neglected.

%\subsection{Monte Carlo calculations}

In order to evaluate the effects of the two-body Coulomb interaction on the strength $m_0$ and
the energy $E_{1,0}$ we have performed a Quantum Monte Carlo (QMC) calculation.
Our recent extension of the Diffusion Monte Carlo (DMC) \cite{Ambr}, in fact,
allows for the projection over the ground state of the system
corresponding to the full Hamiltonian (\ref{ham}).
DMC is a very accurate method to solve the Schroedinger equation, based on projecting out of an
initial trial wavefunction $\Psi_{T}(R)$ the ground state of the Hamiltonian by means
of an importance sampled imaginary time propagation:
\begin{equation}
\Psi_{T}(R)\phi(R,\tau)=\int dR' G(R;R';\tau)\frac{\Psi_{T}(R)}{\Psi_{T}(R')}
\Psi_{T}(R')\phi(R,0)
\end{equation}
In this case $R$ indicates both the space and spin coordinates of the electrons. The Green's Function
for short imaginary time can be approximated by:
\begin{equation}
G(R,R',\Delta\tau)\propto e^{-V_{Rashba}\Delta\tau}e^{-\frac{(R-R')^{2}}{2\frac{\hbar}{m}\Delta\tau}}
e^{-(V_{Coul}\Delta\tau-E_{0})},
\end{equation}
where $V_{Coul}$ is the Coulomb potential, and $E_{0}$ is a constant used to preserve the
normalization. The factor including $V_{Rashba}$ contains an explicit dependence on the spin
coordinates of the electrons. This fact demands a substantial modification of the standard
algorithm, which must now take care of the propagation of the spin degrees of freedom together
with the space coordinates. The expectation giving the quantities of interest can then be evaluated as 
stochastic averages. We should mention that,
as in all applications of standard DMC to many Fermion systems, one must introduce an artificial
constraint (in this case a fixed-phase constraint) in order to avoid the notorious sign problem.
The details of the algorithm and further discussions can be found in Ref. \cite{Ambr}.

\begin{figure}[ht]
\vspace{0.7cm}
\centering
\includegraphics[scale=0.31]{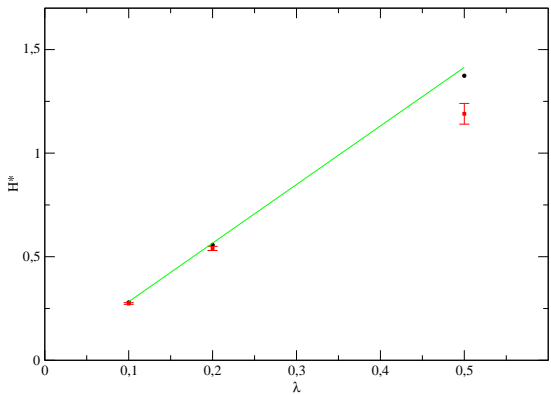}
\caption{(Color online) Average excitation energy estimates are reported in $H^*$ units. The green line and the black dots correspond respectively to the lowest $\lambda$ order estimates (linear) and to the exact values for the independent particle model. The red squares correspond to the QMC estimates in presence of Coulomb interaction.}
\label{fg}
\end{figure}

We have evaluated  the momenta
$m_1$ and $m_0$ by explicitly calculating the expectation values of eqs. (\ref{sumr}) on
the Monte Carlo ground state at $r_s=1$, corresponding to a density $\rho\simeq 3 \cdot 10^{11} cm^{-2}$. 
Calculations were performed using the trial wave function
described in \cite{Ambr}, using the combination of quasi-up and quasi-down single particle states
giving the lowest ground state energy.
The obtained results are reported in Table \ref{tablemc} and Figure \ref{fg} for three
values of $\lambda$ expressed in effective atomic units.
\newline

\begin{table}
\begin{tabular}{cccccc}
\multicolumn{6}{c}{} \\
\hline
\hline
$\lambda$ & $N_-$ & $N_+$ & $m_0$ & $m_1$ & $E_{av}$  \\
\hline
0.1 & 33 & 25 & 2.05(3) & 0.561(3) & 0.274(4)  \\
0.2 & 37 & 21 & 4.20(7) & 2.250(6) & 0.54(1)  \\
0.5 & 48 & 10 & 11.1(5) & 13.19(1) & 1.19(5)  \\
\hline
\hline
\end{tabular}
\caption{Extended DMC ground state estimates for the first and second moments of the structure function and for the 
average 
excitation energy (in $H^*$) for the excitation operator $S_z$ as a function of the SO strength $\lambda$. Beside $\lambda$, 
the number of "quasi-up" ($N_+$) and "quasi-down" ($N_-$) states used in the trial wave function are reported.}
\label{tablemc}
\end{table}

In Fig. 1 we compare the $\lambda$ dependence of the excitation energy
predicted by the independent particle model (with and without corrections at order 
$\lambda^{2}$), and by the DMC calculations. As it can be noticed, the deviations
due to the presence of the electron-electron correlations are rather small. At values
of $\lambda$ of order of the estimates for typical quantum wells they are almost
negligible,  and increase with the strength of the transverse electric field. 
This implies  that the analytical results (\ref{stren}) and 
(\ref{ener}) can be considered very good estimates for the strength and excitation energies
of far infrared absorption of polarized light in the quantum well. This confirms the fact
that a good comparison with experiments would support the physical content of the
2DEG approximation even for as concerns the description of the effects of the well asymmetry.

%\section{Conclusions}

In summary,  we have calculated strength and mean excitation energy of
the transitions induced in a two-dimensional quantum well by a linearly polarized 
electromagnetic wave propagating in the plane of motion of the electrons by means of 
sum-rules techniques. In absence of a transverse electric field there is no absorption of light in the
system. When the electric field is turned on, and the corresponding strength of the
Rashba SO coupling is non-zero,  the predicted strength and mean excitation
energy turn out to be proportional to the Rashba coupling parameter $\lambda$.
Calculations have been performed 
analytically in the single particle picture which neglects the Coulomb interaction and takes
into account the SO potential. The effects of the Coulomb interaction have been accurately
tested by numerical Diffusion Monte Carlo simulations including the recently developed
extension to the treatment of spin-orbit interactions.

%\section{acknowledgements}
We acknowledge useful discussions with S. Gandolfi and L. Mitas.
DMC calculations were performed on the Wiglaf HPC facility of the Physics Department of the University of Trento 
and on the CINECA HPC under a Computing Project Grant of the University of Trento.

\end{document}